\documentclass[prl,superscriptaddress,twocolumn]{revtex4}

\usepackage{amssymb}
\usepackage{natbib}
\usepackage{graphicx}
\usepackage{amsmath}
\usepackage[bookmarks = false]{hyperref}
\usepackage{color}

\begin{document}

\title{Hong-Ou-Mandel Interference between Two Deterministic Collective Excitations in an Atomic Ensemble}

\author{Jun Li}
\affiliation{Hefei National Laboratory for Physical Sciences at Microscale and Department
of Modern Physics, University of Science and Technology of China, Hefei,
Anhui 230026, China}
\affiliation{CAS Center for Excellence and Synergetic Innovation Center in Quantum
Information and Quantum Physics, University of Science and Technology of
China, Hefei, Anhui 230026, China}
\affiliation{CAS-Alibaba Quantum Computing Laboratory, Shanghai 201315, China}
\author{Ming-Ti Zhou}
\affiliation{Hefei National Laboratory for Physical Sciences at Microscale and Department
of Modern Physics, University of Science and Technology of China, Hefei,
Anhui 230026, China}
\affiliation{CAS Center for Excellence and Synergetic Innovation Center in Quantum
Information and Quantum Physics, University of Science and Technology of
China, Hefei, Anhui 230026, China}
\affiliation{CAS-Alibaba Quantum Computing Laboratory, Shanghai 201315, China}
\author{Bo Jing}
\affiliation{Hefei National Laboratory for Physical Sciences at Microscale and Department
of Modern Physics, University of Science and Technology of China, Hefei,
Anhui 230026, China}
\affiliation{CAS Center for Excellence and Synergetic Innovation Center in Quantum
Information and Quantum Physics, University of Science and Technology of
China, Hefei, Anhui 230026, China}
\affiliation{CAS-Alibaba Quantum Computing Laboratory, Shanghai 201315, China}
\author{Xu-Jie Wang}
\affiliation{Hefei National Laboratory for Physical Sciences at Microscale and Department
of Modern Physics, University of Science and Technology of China, Hefei,
Anhui 230026, China}
\affiliation{CAS Center for Excellence and Synergetic Innovation Center in Quantum
Information and Quantum Physics, University of Science and Technology of
China, Hefei, Anhui 230026, China}
\affiliation{CAS-Alibaba Quantum Computing Laboratory, Shanghai 201315, China}
\author{Sheng-Jun Yang}
\affiliation{Hefei National Laboratory for Physical Sciences at Microscale and Department
of Modern Physics, University of Science and Technology of China, Hefei,
Anhui 230026, China}
\affiliation{CAS Center for Excellence and Synergetic Innovation Center in Quantum
Information and Quantum Physics, University of Science and Technology of
China, Hefei, Anhui 230026, China}
\affiliation{CAS-Alibaba Quantum Computing Laboratory, Shanghai 201315, China}
\author{Xiao Jiang}
\affiliation{Hefei National Laboratory for Physical Sciences at Microscale and Department
of Modern Physics, University of Science and Technology of China, Hefei,
Anhui 230026, China}
\affiliation{CAS Center for Excellence and Synergetic Innovation Center in Quantum
Information and Quantum Physics, University of Science and Technology of
China, Hefei, Anhui 230026, China}
\affiliation{CAS-Alibaba Quantum Computing Laboratory, Shanghai 201315, China}
\author{Klaus M\o{}lmer}
\affiliation{Department of Physics and Astronomy, Aarhus University, DK-8000 Aarhus C, Denmark}
\author{Xiao-Hui Bao}
\affiliation{Hefei National Laboratory for Physical Sciences at Microscale and Department
of Modern Physics, University of Science and Technology of China, Hefei,
Anhui 230026, China}
\affiliation{CAS Center for Excellence and Synergetic Innovation Center in Quantum
Information and Quantum Physics, University of Science and Technology of
China, Hefei, Anhui 230026, China}
\affiliation{CAS-Alibaba Quantum Computing Laboratory, Shanghai 201315, China}
\author{Jian-Wei Pan}
\affiliation{Hefei National Laboratory for Physical Sciences at Microscale and Department
of Modern Physics, University of Science and Technology of China, Hefei,
Anhui 230026, China}
\affiliation{CAS Center for Excellence and Synergetic Innovation Center in Quantum
Information and Quantum Physics, University of Science and Technology of
China, Hefei, Anhui 230026, China}
\affiliation{CAS-Alibaba Quantum Computing Laboratory, Shanghai 201315, China}

\begin{abstract}
We demonstrate deterministic generation of two distinct collective excitations in one atomic ensemble, and we realize the Hong-Ou-Mandel interference between them. Using Rydberg blockade we create single collective excitations in two different Zeeman levels, and we use stimulated Raman transitions to perform a beam-splitter operation between the excited atomic modes. By converting the atomic excitations into photons, the two-excitation interference is measured by photon coincidence detection with a visibility of 0.89(6). The Hong-Ou-Mandel interference witnesses an entangled NOON state of the collective atomic excitations, and we demonstrate its two times enhanced sensitivity to a magnetic field compared with a single excitation. Our work implements a minimal instance of Boson sampling and paves the way for further multi-mode and multi-excitation studies with collective excitations of atomic ensembles.
\end{abstract}

\maketitle

When two identical photons are incident on different input ports of a 50:50 beam-splitter, the cases either both photons are transmitted or both are reflected interfere destructively, and coincidence counts become absent between the two output ports. This so-called Hong-Ou-Mandel (HOM) interference~\cite{Hong1987} is a genuine quantum effect with no classical counterpart, and it has been widely used for testing photon indistinguishability~\cite{Santori2002}, generating multi-photon entanglement~\cite{Pan2012} and constructing quantum gates~\cite{Kok2007a} etc. In recent years, HOM interference has been experimentally extended to electrons~\cite{Bocquillon2013}, atoms~\cite{Kaufman2014,Lopes2015}, phonons~\cite{Toyoda2015} and surface plasmons~\cite{Heeres2013,Fakonas2014}. For bosons it is associated with bunching, while for fermions it leads to antibunching.

Atomic ensembles are excellent media for photon  storage~\cite{Hammerer2010,Sangouard2011,Bussieres2013,Heshami2016} since the atom-light interaction is collectively enhanced, and the collective excitations have bosonic character. When supplemented with the Rydberg blockade non-linearity~\cite{Lukin2001,Saffman2010c,Pritchard2010,Dudin2012,Peyronel2012a,Firstenberg2013a,Hofmann2013a,Ebert2013} and multilevel collective encoding~\cite{Brion2007,Saffman2008,Nielsen2010}, an atomic ensemble is able to encode many qubits and becomes a promising platform for quantum computing and quantum simulation. In this paper we report deterministic generation of two distinct collective excitations which reside in the same atomic ensemble, and realize the HOM interference between them. Our work implements a minimal instance of Boson sampling and it brings promises to use and manipulate more collective excitations in one atomic ensemble for advanced quantum applications in a collective way.

\begin{figure}[h]
\includegraphics[width=\columnwidth]{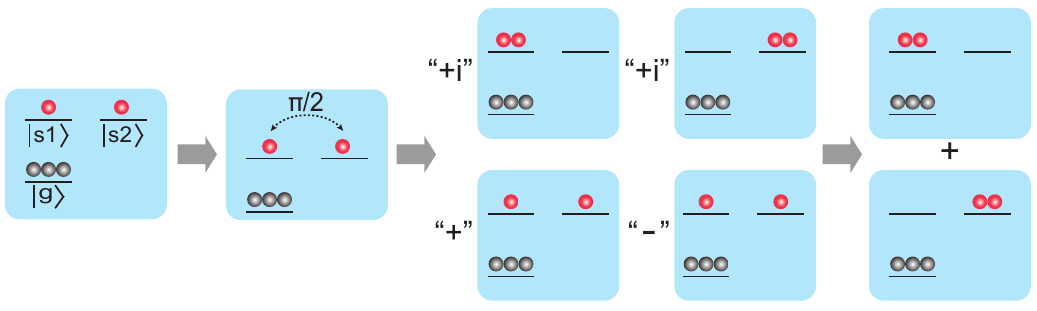}
\caption{Conceptual diagram of HOM interference with collective excitations. First, a collective two-excitation state is generated in a single atomic ensemble. Afterwards, a $\pi/2$ Raman pulse coupling $\left|s1\right>$ and $\left|s2\right>$ applies a ``beam-splitter'' operation for the collective excitations, coupling $\left|1,1\right>_{S1,S2}$ to $\left|2,0\right>_{S1,S2}$ and $\left|0,2\right>_{S1,S2}$. The amplitudes contributing to the remaining state $\left|1,1\right>_{S1,S2}$ occupation interfere destructively and, as a consequence, a NOON state ($N=2$) is created.}\label{scheme}
\end{figure}

Our experimental scheme is depicted conceptually in Fig.~\ref{scheme}. Initially all atoms are prepared in a ground state $|g\rangle$. We create collective excitations in a deterministic fashion by making use of Rydberg blockade~\cite{Lukin2001,Saffman2010c}. Thus, a single excitation can be collectively transferred, via a Rydberg state, to a different ground state $|s1\rangle$, yielding the collective state,  $|1\rangle_{S1}=N^{-1/2}\sum_{j}^{N}|g...(s1)_j...g\rangle$, in which $j=1...N$ labels the different atoms in the ensemble. Alternatively, we can create the excitation in another ground state $|s2\rangle$ to produce the collective state $|1\rangle_{S2}= N^{-1/2}\sum_{j}^{N}|g...(s2)_j...g\rangle$. Note that stimulated Raman light pulses coupling  $\left|s1\right>$ and $\left|s2\right>$ coherently couple the collective states $|1\rangle_{S1}$ and $|1\rangle_{S2}$ and serve as a beam-splitter for the collective excitation shared between them. When a $\pi/2$ Raman pulse is applied, $|1\rangle_{S1}$ will thus be converted into the superposition state $1/\sqrt{2}(|1\rangle_{S1}+i|1\rangle_{S2})$, and $|1\rangle_{S2}$ will be converted to $1/\sqrt{2}(|1\rangle_{S2}+i|1\rangle_{S1})$.

\begin{figure}[h]
\includegraphics[width=\columnwidth]{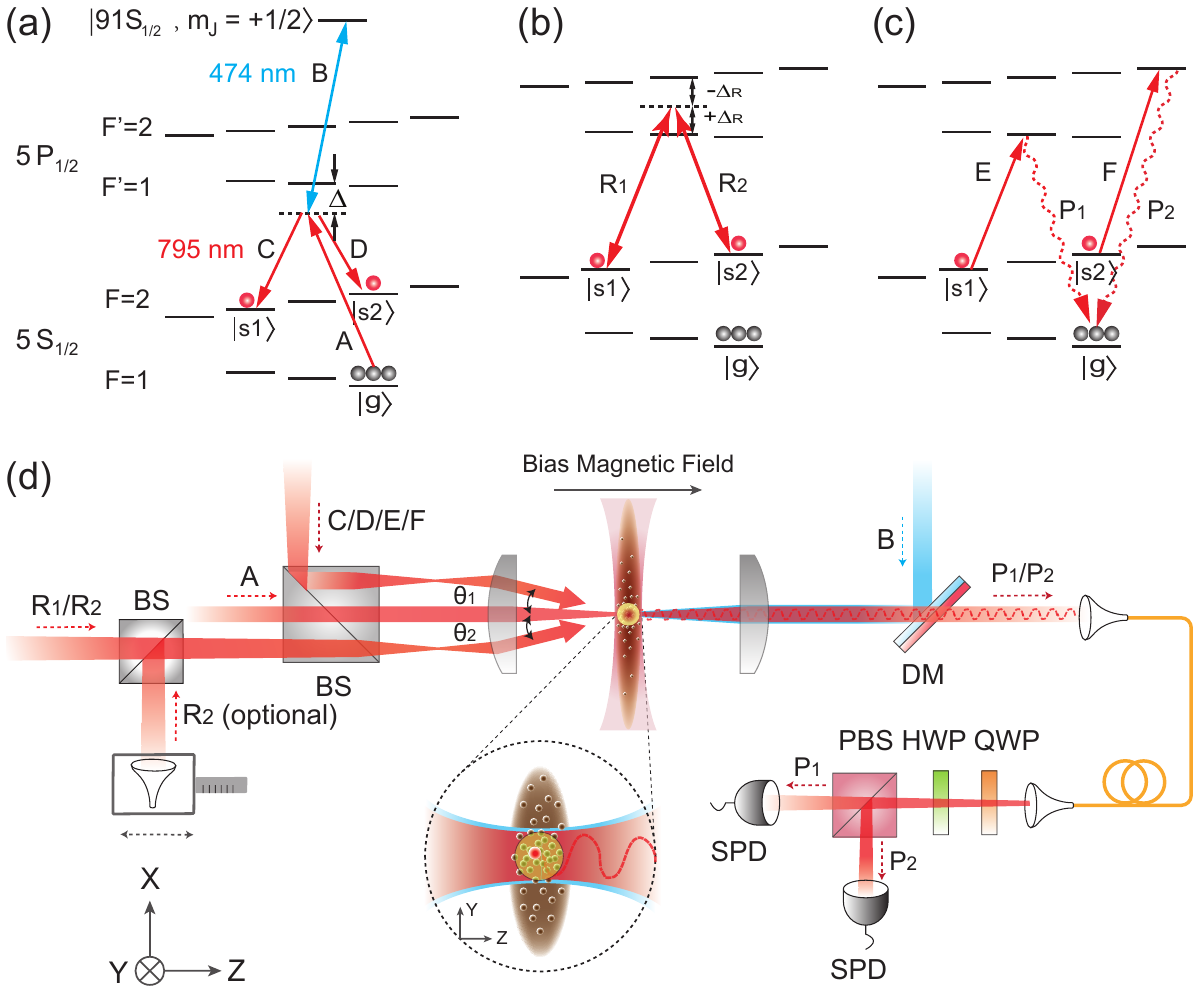}
\caption{Level scheme and experimental layout. (a) Two collective excitations $|1\rangle_{S1}$ and $|1\rangle_{S2}$ are generated sequentially via Rydberg blockade ($AB\rightarrow BC\rightarrow AB\rightarrow BD$). The single-photon detuning of the Rydberg excitation process is set to $\Delta=-40$ MHz relative to the intermediate state $|5P_{1/2},F^{'}=1,m_{F^{'}}=0\rangle$. (b) A beam-splitter operation between the collective occupation of states $|s1\rangle$ and $|s2\rangle$ is realized via stimulated Raman transition. R$_{1}$ and R$_{2}$ are the Raman light fields and $|\pm\Delta_{R}|\approx$ 407 MHz. (c) To convert collective atomic excitations to single photons for detection, we sequentially apply read control fields ($E\rightarrow F$), resonant with different upper levels. A collective atomic excitation will return to the initial state $|g\rangle$ emitting a single phase matched photon due to collective interference. (d) Experimental layout. Control beams are combined with beam-splitters (BS) and dichroic mirrors (DM). One optional port for R$_{2}$ is used for observing the HOM dip. At the atom location, the beam waist is $\sim7~\mu$m for A/B, $55~\mu$m for C/D/E/F, $180~\mu$m for R$_{1}$/R$_{2}$. The small angles between the beams are $\theta_1$=6$\,^{\circ}$ and $\theta_2$=3$\,^{\circ}$. The photons converted from the collective excitations (P1 and P2) are emitted in the phase-matching direction ($Z$), and collected with a single-mode fiber and detected subsequently with two single-photon detectors (SPD). HWP denotes half-wave plate, QWP denotes quarter-wave plate, and PBS denotes polarizing beam-splitter.}\label{setup}
\end{figure}

If $N$ is large, the ground state $|g\rangle$ is not appreciably depleted by the transfer of one or two atoms from $|g\rangle$ to the other ground states, and by collectively transferring one atom to $\left|s1\right>$ and another one to $\left|s2\right>$, we thus produce the two-excitation state, $|1,1\rangle_{S1,S2} \simeq N^{-1}\sum_{jk}^{N}|g...(s1)_j...(s2)_k...g\rangle$. The collective excitations are well described as independent harmonic oscillator degrees of freedom, and hence the beam splitter action implemented by the Raman pulses on $|1,1\rangle_{S1,S2}$ is expected to yield output states $|1,1\rangle_{S1,S2}$, $|2,0\rangle_{S1,S2}$ and $|0,2\rangle_{S1,S2}$, where the last two terms denote collective states with two atoms occupying the internal states $|s1\rangle$ and $|s2\rangle$, respectively. After the $\pi/2$ Raman pulse, the state $|1,1\rangle_{S1,S2}$ with one excitation in each collective mode is expected to disappear due to the HOM interference.

We make use of a small atomic ensemble of $^{87}\rm{Rb}$ atoms loaded from a magneto-optical trap (MOT) and confined in an optical dipole trap (852~nm). The aforementioned states $|g\rangle$, $|s1\rangle$ and $|s2\rangle$ correspond to different hyperfine Zeeman sublevels as shown in Fig.~\ref{setup}a. Initially all atoms are prepared in $|g\rangle$ via optical pumping. Rydberg excitation is realized via a two-photon excitation process with the tightly focused manipulation beams A and B in Fig.~\ref{setup}a. We first excite one atom to the $\left|91S_{1/2},m_{J}=+1/2\right>$ Rydberg state~\cite{Dudin2012e} and afterwards drive it back to $|s1\rangle$. In this way, the collective state $|1\rangle_{S1}$ is created. Repeating a similar procedure, without disturbing the $\left|s1\right>$ ground state because of the different light polarizations and Zeeman energy level shifts, we collectively transfer a second atom to $|s2\rangle$ and thus we prepare the collective two-excitation state $|1,1\rangle_{S1,S2}$. Then, we employ stimulated Raman light pulses between $|s1\rangle$ and $|s2\rangle$, see Fig.~\ref{setup}b, to perform the HOM beam-splitter operation. Finally, we detect the two-excitation states by converting the collective occupation of the atomic states $\left|s1\right>$ and $\left|s2\right>$ sequentially into single photons as shown in Fig.~\ref{setup}c and collecting these with single-mode fibers at the phase-matching direction. The geometric layout for our setup is shown in Fig.~\ref{setup}d, and more experimental details are given in Supplemental Material.

\begin{figure}[h]
\includegraphics[width=.85\columnwidth]{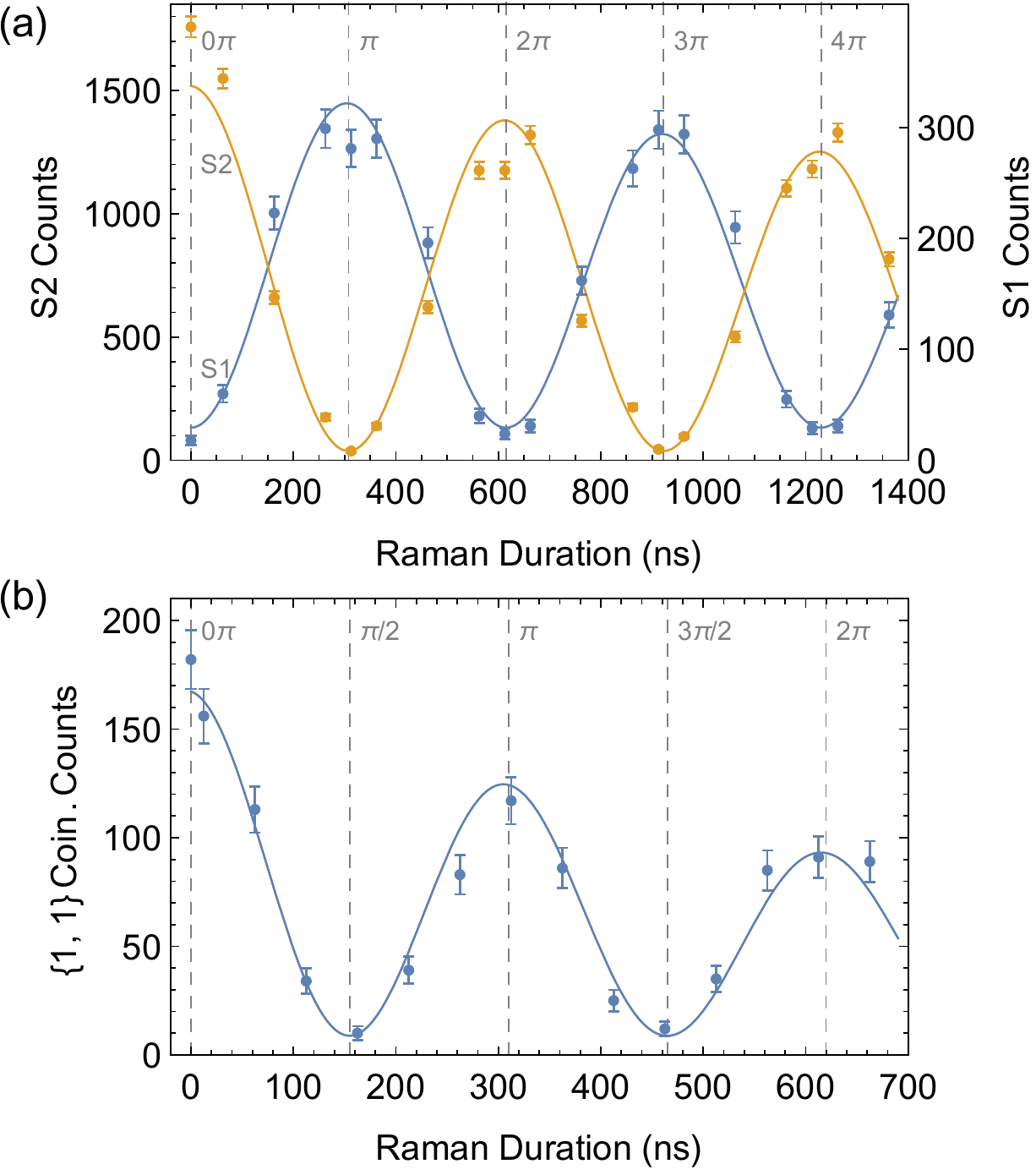}
\caption{Rabi oscillation with one or two collective excitations. (a) A single collective excitation $\left|1\right>_{S2}$ is prepared, and the read-out signal oscillates as a function of the Raman pulse duration. Counts of photons from $\left|1\right>_{S1}$ is relatively small due to low excitation-to-photon conversion efficiency. We fit the data with a damped oscillation function, which calibrates the Raman Rabi frequency to be $2\pi\times 1.626(5)$~MHz. (b) When both collective modes are excited in the state $\left|1,1\right>_{S1,S2}$, the (\{1, 1\}) photon coincidence signal also oscillates as a function of Raman pulse durations. We fit the data with a damped oscillation function, which results in an oscillation frequency of $2\pi\times 3.23(4)$~MHz and coincidence dips at durations of a $\pi/2$ and a $3\pi/2$ pulse. In both plots the error bars indicate $\pm1$~s.d.}\label{rabi}
\end{figure}

We first characterize the quality of the collective excitations and the Raman beam-splitter operation. It is a prerequisite to study HOM interference that the collective two-excitation state $|1,1\rangle_{S1,S2}$ contains precisely one excitation in each of the collective modes populating the sublevels $\left|s1\right>$ and
$\left|s2\right>$. In order to characterize our state preparation procedure, we prepare the singly excited states $\left|1\right>_{S1}$ and $\left|1\right>_{S2}$, and we measure the second-order autocorrelation $g^{(2)}$ for the photons converted from these states~\cite{Dudin2012}. The results are $g^{(2)}=0.046(13)$ for $\left|1\right>_{S1}$ and $g^{(2)}=0.062(14)$ for $\left|1\right>_{S2}$, which confirms that high-quality single collective excitations are indeed prepared. We have also examined the stimulated Raman process. With the singly excited $|1\rangle_{S2}$, we apply the Raman light with different pulse durations and observe the Rabi oscillation between the collective states $|1\rangle_{S1}$ and  $|1\rangle_{S2}$. As shown in Fig.~\ref{rabi}a, the photon counts detected from these states oscillate with the same period, indicating that the collective excitation evolves coherently between $|1\rangle_{S1}$ and  $|1\rangle_{S2}$ with a high fidelity. We attribute the slow exponential decay of the signal to photon scattering and experimental imperfections such as impure light polarizations, inhomogeneous intensity and mode divergence of the Raman beams.

\begin{figure}[h]
\includegraphics[width=.7\columnwidth]{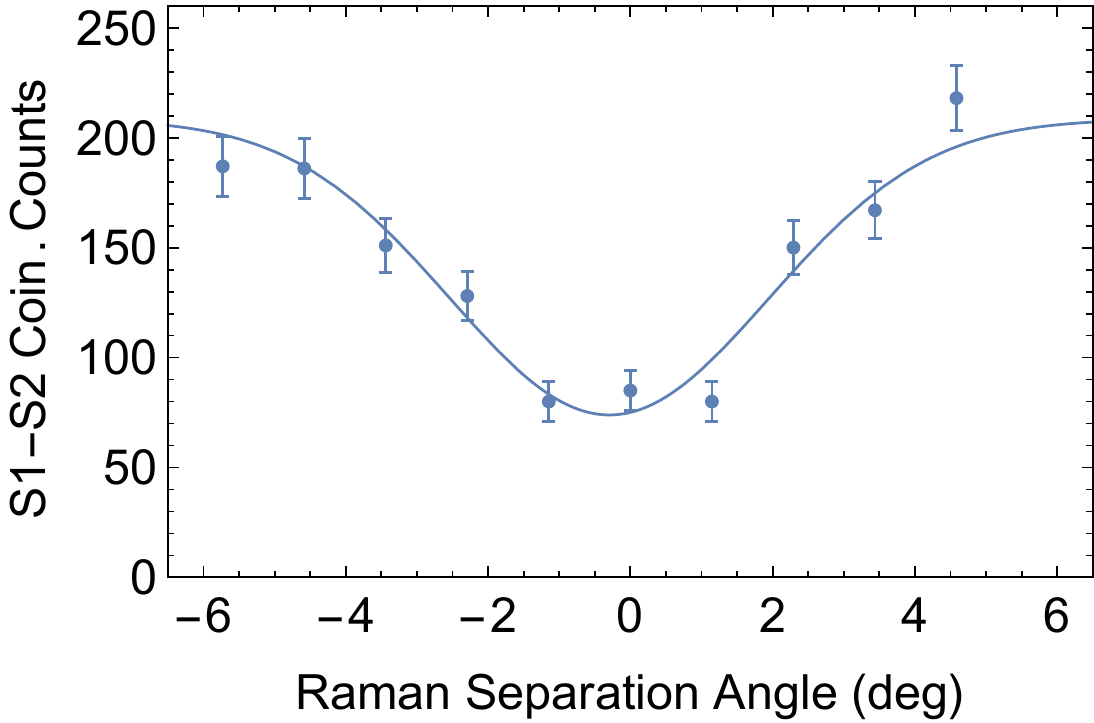}
\caption{Measurement of HOM dip. By tuning the separation angle between the Raman beams, we continuously modify the distinguishability between the collective excitations after beam-splitting. At the position of $\theta\simeq0$, the best indistinguishability is achieved and a HOM dip is observed. For each angle, the beam-splitting Raman pulse ($\pi/2$) is verified to give a fidelity better than 96\%. The error bars indicate $\pm1$~s.d.}\label{homdip}
\end{figure}

Subsequently, we prepare the collective two-excitation state $|1,1\rangle_{S1,S2}$ and we record coincidence counts between the photons converted from the atomic excitations. The result is shown in Fig.~\ref{rabi}b. By changing the Raman pulse duration before the photon emission, we effectively vary the beam-splitting ratio between the collective atomic modes. At $t=155$~ns, which corresponds to a $\pi/2$ pulse with 50:50 splitting ratio, we observe the smallest value of the coincidence rate. This is a strong indication of HOM interference. By comparing the coincidence rate between $t=155$~ns and $t=0$~ns, we get a two-excitation interference visibility of $V=1-2~\rm{C(155ns)/C(0ns)}=0.89(6)$, which is significantly higher than the classical limit of 0.5~\cite{Ou1989}. The non-zero counts at the coincidence minimum primarily arise from dark counts of the single-photon detectors and crosstalk between the two channels during the read-out process.

In a typical HOM interference experiment, by tuning particles from indistinguishable to distinguishable, the coincidence dip will disappear. In our previous measurement in Fig.~\ref{rabi}b, the Raman beams are co-propagating in the same spacial mode and thus the collective excitations after Raman manipulation share the same wave vector $\bf{k}$ and interfere perfectly. By applying a small angle between the Raman beams, we can induce a wave vector difference between the Raman coupled collective states. \textit{I.e.}, the collective excitations evolve under the Raman $\pi/2$ pulse as
\begin{align*}
\left|1,\textbf{k}\right>_{S1}\to\dfrac{1}{\sqrt{2}}(\left|1,\textbf{k}\right>_{S1}+i\left|1,\textbf{k}+\Delta\textbf{k}\right>_{S2})
\\
\left|1,\textbf{k}\right>_{S2}\to\dfrac{1}{\sqrt{2}}(\left|1,\textbf{k}\right>_{S2}+i\left|1,\textbf{k}-\Delta\textbf{k}\right>_{S1})
\end{align*}
where $\Delta\textbf{k}=\textbf{k}_{\rm R1}-\textbf{k}_{\rm R2}$, $\textbf{k}_{\rm R1}$ and $\textbf{k}_{\rm R2}$ denote the wave vectors of the two Raman fields. As shown in Fig.~\ref{setup}d, fine tuning of the separation angle is realized by feeding one Raman beam through an optional port and displacing it with a translation stage. Experimental results for the measured coincidences as functions of the Raman separation angle is shown in Fig.~\ref{homdip}. The raw visibility is 0.65(11), which is much lower than the result in Fig.~\ref{rabi}b. This is due to the wave vector dependence of the detection efficiency since a single-mode fiber is used to collect the excitation-converted photons. When the Raman separation angle is large compared with the photon collection angle, the excitations $|1,\textbf{k}+\Delta\textbf{k}\rangle_{S2}$ and $|1,\textbf{k}-\Delta\textbf{k}\rangle_{S1}$ will no longer be efficiently detected, which thus lowers the baseline of Fig.~\ref{homdip} by a factor of 2. Taking this deficiency into account, we infer a corrected visibility of $V_{\rm{cor}}=0.82(5)$, in agreement with the result in Fig.~\ref{rabi}b.

\begin{figure}[h]
\includegraphics[width=.85\columnwidth]{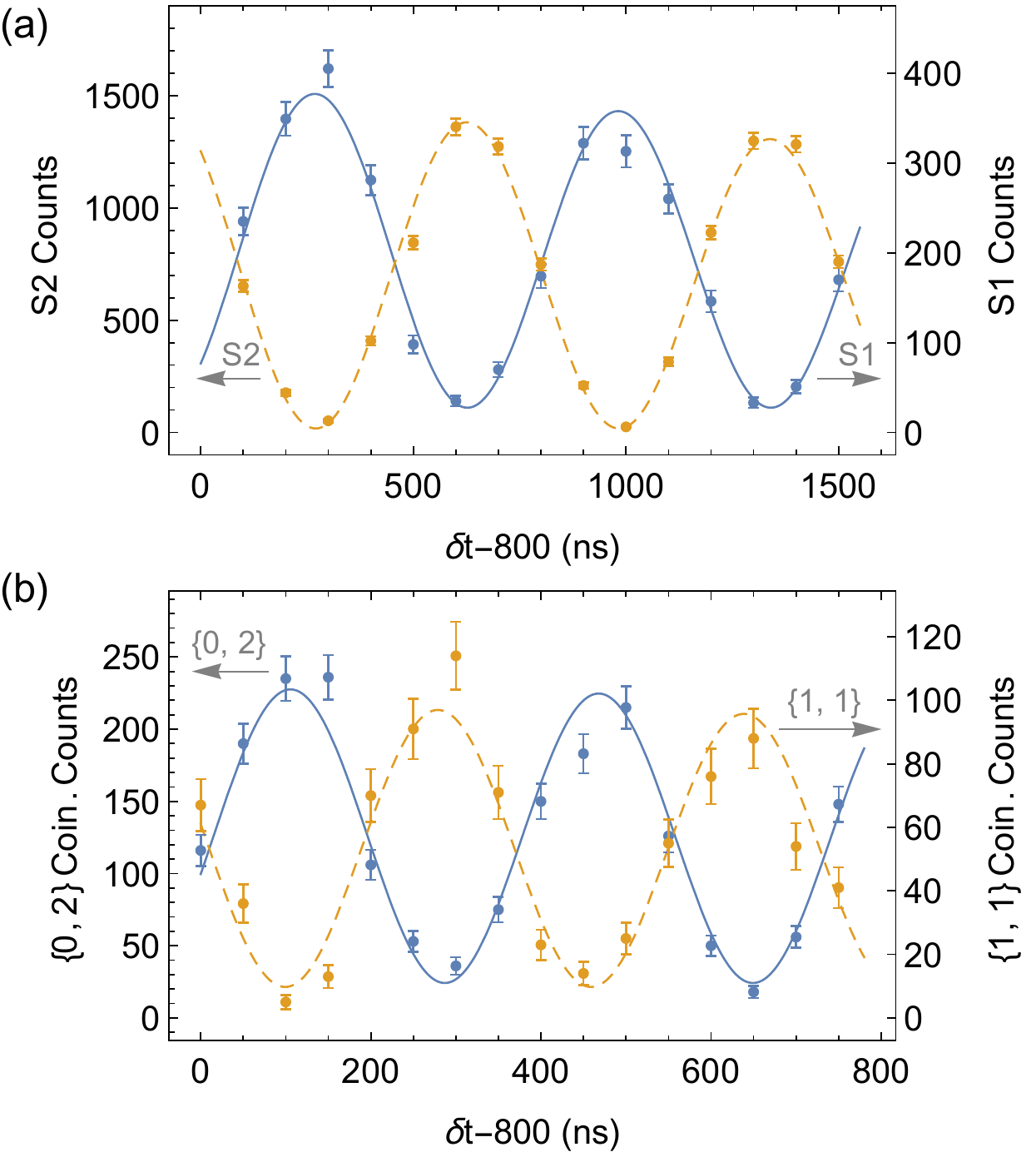}
\caption{Ramsey interference with NOON states of collective excitations. (a)
A single excitation of $\left|1\right>_{S1}$ is prepared. The read-out photons oscillate as function of the time interval between the two $\pi/2$ pulses with a Larmor precession frequency of 1.401(1)~MHz, which is consistent with the applied magnetic field. (b) The collective two-excitation state $\left|1,1\right>_{S1,S2}$ is prepared and produces a NOON state with $N=2$ after the first $\pi/2$ pulse. \{0, 2\} and \{1, 1\} coincidence measurement are performed, where \{0, 2\} denotes two photons from the atomic state  $|s2\rangle$ and \{1, 1\} denotes one photon from each of the two states $|s1\rangle$ and $|s2\rangle$. The oscillation frequency of the coincidence signals is two times faster than the intensity signal modulation from the single excited states. Due to a restriction by our instrumental setup, we can only make these measurements starting from $\delta t\geq800$~ns, and the time of all our data points is shifted by $800$~ns. In both plots the error bars indicate $\pm1$~s.d.}\label{noon}
 \end{figure}

The HOM interference results witness the preparation of entangled NOON state $1/{\sqrt{2}}~(\left|2,0\right>_{S1,S2}+e^{i\phi_{2}(t)}\left|0,2\right>_{S1,S2})$ without employing atom-atom interaction, which is useful for quantum-enhanced phase measurements~\cite{Chen2010e}. In our setup, the Zeeman splitting induces a time-dependent phase shift of $\phi_1(t)=\lambda B\,t$ between $|1\rangle_{S1}$ and $|1\rangle_{S2}$, where $B$ is the magnetic field and $\lambda$ is a coefficient representing the different magnetic moments of the atomic states. A NOON state with $N=2$ features components with a phase difference evolving at twice the rate, \textit{i.e.} $\phi_{2}(t)=2\lambda B\,t$. As a reference, we first prepare the singly excited state $\left|1\right>_{S1}$, and we employ a Ramsey interferometer configuration: a $\pi/2$ Raman pulse, a time interval $\delta t$ for free phase evolution, a second $\pi/2$ Raman pulse, and a read-out of the $|s2\rangle$ population. The oscillations at 1.401(1)~MHz, shown in Fig.~\ref{noon}a, are consistent with the Zeeman energy difference between $\left|s1\right>$ and $\left|s2\right>$ induced by a 1 Gauss external magnetic field. Next, we prepare $\left|1,1\right>_{S1,S2}$ and we apply the same Ramsey interferometer procedure. In principle, to observe super-resolving oscillations, we can record either the \{1, 1\} or the \{0, 2\} coincidences, where \{i, j\} denote the photon counts originating from the  collective occupation of $\left|s1\right>$ and $\left|s2\right>$. Fig.~\ref{noon}b shows the super-resolving oscillation of the coincidence counts with a frequency of 2.77(1)~MHz, resulting from the doubled collective phase evolution speed. The \{0, 2\} oscillation visibility is measured to be 0.81(3), which is far beyond the classical bound of 1/3~\cite{Afek2010} and implies that we are able to measure the magnetic field with genuine quantum enhancement.

In summary, we have experimentally realized the HOM interference between two collective excitations in an atomic ensemble quantum memory. Our work implements a minimal instance of Boson sampling and paves the way for collectively encoding and manipulating more excitations in one atomic ensemble for quantum-enhanced applications. The current experiment is limited by the low overall detection efficiency. By employing cavity enhancement to boost excitation-to-photon conversion efficiency~\cite{Simon2007b,Bao2012g} or using efficient state-selective ionizing detection~\cite{Henkel2010}, multiple-excitation coincidence rate will get much higher and experiments with up to 7 collectively excitated modes may become possible with the current setup. An atom species with a large electronic angular momentum and a large nuclear spin such as $^{165}$Ho ~\cite{Saffman2008} may provide many more hyperfine ground states for collective excitations and, e.g., permit implementation of Boson sampling beyond reach of current classical computational methods~\cite{Wu2016}.

This work was supported by the National Natural Science Foundation of China, the Chinese Academy of Sciences, and the Ministry of Science and Technology of China. KM acknowledges support from the Villum Foundation. X.-H.B. acknowledge special support from the Youth Qianren Program.


\bibliography{myref}

\section*{Supplemental Material}

\textbf{Technical details.}
The ensemble in the dipole trap has a spatial extent of 250 $\mu$m, 40 $\mu$m and 13 $\mu$m in the direction of $X$, $Y$ and $Z$ respectively. The Rydberg manipulation beams are tightly focused to a waist of 7 $\rm{\mu m}$, selecting a region of interest that contains about 300 atoms. A magnetic field of 1 Gauss is applied in the $Z$ direction to lift the Zeeman degeneracy. We set the frequencies of the two Raman beams such that the single-photon detuning is $|\Delta_{R}|\approx 407$  MHz with opposite signs for  $\left|5P_{1/2},F^{'}=1,m_{F^{'}}=0\right>$ and $\left|5P_{1/2},F^{'}=2,m_{F^{'}}=0\right>$  (shown in Fig.~\ref{setup}b). They couple $\left|s1\right>$ and $\left|s2\right>$ efficiently while the intermediate states are eliminated adiabatically. For the detection of $\left|1\right>_{S1}$, the phase-matching condition requires: $\textbf{k}_{\rm A}+\textbf{k}_{\rm B}-\textbf{k}_{\rm B}-\textbf{k}_{\rm C}+\textbf{k}_{\rm E}-\textbf{k}_{\rm P1}=0$ . Here $\textbf{k}_{\rm i}$ (i=A, B, ...) denotes the corresponding wave vector of the light field. The difference between $\textbf{k}_{\rm C}$ and $\textbf{k}_{\rm E}$ is negligible. Hence, $\textbf{k}_{\rm P1}$= $\textbf{k}_{\rm A}$. It means the read-out photons will be emitted in the same direction as A. With a similar analysis, we find out that the $\left|1\right>_{S2}$ read-out photons will also be emitted in the same direction. The overall detection efficiency is 0.3\% for $\left|1\right>_{S1}$ and 1.2\% for $\left|1\right>_{S2}$. During the evaluation of $g^{(2)}$ for $\left|1\right>_{S2}$, we measure the read-out photons directly. While for the evaluation of $g^{(2)}$ for $\left|1\right>_{S1}$, we transfer $\left|1\right>_{S1}$  into $\left|1\right>_{S2}$ first with a Raman $\pi$ pulse and measure the $|1\rangle_{S2}$ read-out photons instead.

\textbf{Time sequences.}
Our experiment runs with a repetition rate of 8~Hz. In each experimental cycle, the atoms are loaded into a MOT with a duration of 100 ms. Afterwards a 6 ms temporal dark-MOT phase and a 3 ms molasses cooling phase are applied sequentially. Later the cooling beam is turned off and the atoms outside of the dipole trap region are allowed to fall down under gravity for 12 ms. The small atomic ensemble confined in the dipole trap has a temperature of 12 $\mu$K. Then we begin to repeat the experimental trials for 1000 times before next MOT loading. Each trial lasts 10.7$\mu$s. In each trial, we initialize the atoms into the state $|g\rangle$ and apply the manipulation pulses in Fig.~\ref{setup}. The duration is 200 ns for $|g\rangle$-to-Rydberg excitation, 500 ns for Rydberg-to-$|1\rangle_{S}$ transfer. 100 ns intervals are inserted between the pulses to eliminate crosstalk. At the end of each trial, we apply the read pulses for sequential photonic detection of the collective excitations. Note that, residual Rydberg excitations are cleaned in between the preparations of $|1\rangle_{S1}$ and $|1\rangle_{S2}$. Besides, the dipole trap is temporarily shut off to avoid differential light shifts during Rydberg state manipulations.

\textbf{Preparation efficiency.}
In principle, preparation of collective excitations by Rydberg blockade is deterministic, but experimental imperfections (e.g. a spatially inhomogeneous beam intensity) reduces the preparation efficiency. The $|g\rangle$-to-Rydberg transfer efficiency is inferred to be around $\eta_r=85\%$, by evaluating the damped Rabi oscillation between $|g\rangle$ and the Rydberg state. The Rydberg-to-$|s\rangle$ transfer efficiency is inferred to be $\eta_s=65\%$, by comparing the signal intensity of ground state ($|s1\rangle$ or $|s2\rangle$) retrieval with the signal intensity of Rydberg retrieval. Therefore, the overall preparation efficiency for a single excitation $|1\rangle_{S1}$ or $|1\rangle_{S2}$ is estimated to be around $\eta_r\eta_s=55\%$, and for dual excitations $|1,1\rangle_{S1,S2}$, it is around $\eta_r^2\eta_s^2=30\%$. These values significantly outperform the traditional method of probabilistic preparation via Raman scattering~\cite{Chen2010e}, which succeeds with a probability of $\sim10^{-3}$ for a single excitation and $\sim10^{-5}$ for dual excitations.

\textbf{Memory Lifetime.}
We measure the $1/e$ lifetime of the ground-state excitations by generating $|1,1\rangle_{S1,S2}$ and collecting the converted single photons in the phase-matching direction after a variable storage time. The results are 29(3) $\mu$s for $|1\rangle_{S1}$ and 29(2) $\mu$s for $|1\rangle_{S2}$. They are sufficiently long compared with the preparation and manipulation timescale of $1\sim2$ $\mu$s in the experiment. The current lifetime is mainly limited by motional dephasing. By employing motional confinement in the X direction with optical lattice, the lifetime can be significantly prolonged for more advanced applications.


\end{document}